\documentclass[aps,prb,superscriptaddress,reprint,floatfix]{revtex4-2}
\usepackage{graphicx}
\usepackage{dcolumn}
\usepackage{multirow}
\usepackage{amsfonts}
\usepackage{amsmath}
\usepackage{amssymb}
\usepackage{newtxmath}
\usepackage{dcolumn}% Align table columns on decimal point
\usepackage{subfigure}
\usepackage{braket}
\usepackage{color}
\usepackage{booktabs}
\usepackage{colortbl}
\usepackage{rotating}

\usepackage{multirow}
\usepackage{epstopdf}
\usepackage[colorlinks=true,linkcolor=blue,anchorcolor=blue,filecolor=blue,urlcolor=blue,citecolor=blue]{hyperref}
\usepackage[normalem]{ulem}
\usepackage{natbib}

\graphicspath{{graphs/}}

\begin{document}

\title{Nuclear quantum effects on the quasiparticle properties of the chloride anion aqueous solution within the GW approximation}
\author{Fujie Tang} \affiliation{Department of Physics, Temple University, Philadelphia, PA 19122, USA}
\author{Jianhang Xu} \affiliation{Department of Physics, Temple University, Philadelphia, PA 19122, USA}
\author{Diana Y. Qiu}
\thanks{Corresponding author. Email:diana.qiu@yale.edu}
\affiliation{Department of Mechanical Engineering and Materials Science, Yale University, New Haven, CT 06511}
\author{Xifan Wu}
\thanks{Corresponding author. Email:xifanwu@temple.edu}
\affiliation{Department of Physics, Temple University, Philadelphia, PA 19122, USA}
\affiliation{Institute for Computational Molecular Science, Temple University, Philadelphia, PA 19122}
\date{\today}

\begin{abstract}
\par Photoelectron spectroscopy experiments in ionic solutions reveal important electronic structure information, in which the interaction between hydrated ions and water solvent can be inferred. Based on many-body perturbation theory with GW approximation, we theoretically compute the quasiparticle electronic structure of chloride anion solution, which is modeled by path-integral $ab$ $initio$ molecular dynamics simulation by taking account the nuclear quantum effects (NQEs). The electronic levels of hydrated anion as well as water are determined and compared to the recent experimental photoelectron spectra. It is found that NQEs improve the agreement between theoretical prediction and experiment because NQEs effectively weaken the hybridization of the between the $\rm Cl^-$ anion and water. Our study indicates that NQEs plays a small but non-negligible role in predicting the electronic structure of the aqueous solvation of ions of the Hofmeister series.
 \end{abstract}

\maketitle

\section{INTRODUCTION}

\par Ionic solutions are ubiquitous in nature. Among them, the solvated hydrated chloride ($\rm Cl^-$) anions  play an important role in numerous biochemical~\cite{Bezanilla2008}, chemical~\cite{Knipping2000,Spicer1998,Foster2001}, and geological processes~\cite{Beekman2011}. For example, the $\rm Cl^-$ anion, as a member of the Hofmeister series of ions, impacts the solubility of proteins in water~\cite{Hofmeister1888,Xie2013,Zhang2006a}. As another example, $\rm Cl^-$ ion channels have important functionalities in the cell membrane, which allow for the passage of ions from one side of the membrane to the other~\cite{Dalemans1991,Gaiduk2016}. Not surprisingly, the precise picture of the interaction between hydrated $\rm Cl^-$ and the hydrogen bond network of liquid water continues to be at the center of scientific interest, by joint efforts from both theory and experiment~\cite{Cummings1980,Ohtaki1993,Leberman1995,Ghosal2005,Marcus2009,Piatkowski2014}.

\par The arrangement of water molecules surrounding the hydrated $\rm Cl^-$ ions can be directly detected by neutron scattering or X-ray diffraction~\cite{Cummings1980,Copestake1985,Yamagami1995,Megyes2008} as well as X-ray absorption (XAS)~\cite{Tongraar2010,Dang2006} experiments. Complementary to the scattering experiments, the photoelectron spectroscopy (PES) has recently emerged as an important experimental technique~\cite{Gaiduk2016,Delahay1982,Winter2005,Winter2006,Seidel2011}, in which the electronic structure of aqueous solution is probed. During the process of PES, a valence electron is excited into the vacuum by absorbing energy of an incident photon. Based on the difference between the excitation photon energy and the kinetic energy of the emitted electron, the electron binding energies of both solvent and solvated ions can be determined. In a recent PES experiment by using the advanced microjet technique~\cite{Gaiduk2016}, it was revealed that the valence $3p$ band of the solvated $\rm Cl^-$ anion is about 1.71 eV above the $\rm H_{2}O$ $\rm 1b_1$ band. The energetics of the solvated $\rm Cl^-$ relative to those of liquid water provides an important information on the ion-water interactions in terms of electronic structure.

\par The electronic structure of ionic solutions can also be predicted by first-principles calculations, in which both accurate modeling of electronic structure and the solvation structure are required. Density functional theory (DFT) has been conventionally applied to compute the ionization energies in aqueous solutions due to its low computational cost. However, the electronic structure predicted by DFT overestimates the charge transfer~\cite{Cohen2008,Cohen2012}, which also tends to overestimate the anion and water interactions. Therefore, the predicted energetics of hydrated $\rm Cl^-$ anion was found to be lower than the experimental measurements as a general trend~\cite{Zhang2013a,Bankura2015,Gaiduk2016,DelloStritto2020}. The above predicted energetics of hydrated $\rm Cl^-$ can be improved if hybrid DFT was used instead of the semi-local exchange-correlation (XC) approximations~\cite{Zhang2013a,Bankura2015,Gaiduk2016,DelloStritto2020}. On the other hand, DFT as implemented in its current formalism~\cite{Cohen2012,Cohen2008}, is a ground state theory; the PES experiments however involve single-particle excitations. Therefore, its theoretical modeling demands the electronic excitation theory, in which the electronic screening on the quasiparticle should be properly treated~\cite{Hedin1970,Hybertsen1986,Swartz2013,Chen2016,Gaiduk2016}. Very recently, the many-body perturbation theory such as the GW approximation has been successfully applied in the ionic solutions~\cite{Gaiduk2016}, which yields largely improved energetics for both liquid water and solvated ions in water.

\begin{table}[htbp]
		\vspace{-0.7em}
	\centering
	\caption{The energy separation $\Delta E_{p}$ = $E(\rm H_{2}O(1b_1))$ - $E({\rm Cl}^-(3p))$ and $\Delta E_{s}$ = $E(\rm H_{2}O(2a_1))$ - $E({\rm Cl}^-(3s))$ of the qDOS of a 0.87 M $\rm Cl^-$ solution computed using DFT (PBE, SCAN) and the many-body perturbation theory at the static COHSEX and GPP levels of the theory. The units are eV. The error bar for energy levels are $\sim$0.10 eV. The experimental data are taken from the peak positions of the photoelectron spectrum in the Ref.~\cite{Gaiduk2016}.}

\setlength{\tabcolsep}{2.5mm}
	\begin{tabular}{cccc}
		\toprule
		&       &$\Delta E_{s}$ (eV)&$\Delta E_{p}$ (eV)\\
		\midrule
		\multirow{4}[8]{*}{AIMD} & PBE   &6.56 & 0.67 \\
		\cmidrule{2-4}          & SCAN  & 7.05&0.74 \\
		\cmidrule{2-4}          & G$_0$W$_0$@COHSEX &8.35 & 1.00 \\
		\cmidrule{2-4}          & G$_0$W$_0$@GPP & 8.10& 0.85 \\
		\midrule
		\multirow{4}[8]{*}{PI-AIMD} & PBE  &6.65 & 0.90 \\
		\cmidrule{2-4}          & SCAN & 7.14& 1.00 \\
		\cmidrule{2-4}          & G$_0$W$_0$@COHSEX &8.72 & 1.52 \\
		\cmidrule{2-4}          & G$_0$W$_0$@GPP &8.41 & 1.25 \\
		\midrule
		\multicolumn{2}{c}{Experimental} & 9.9$\sim$11.7& 1.71 \\

		\bottomrule
	
	\end{tabular}%
	\label{tab:deltaE}%
		\vspace{-0.8em}
\end{table}%

\par As far as the modeling of solvation structure is concerned, {\it ab initio} molecular dynamics (AIMD)~\cite{marx2009ab,car1985} simulation has provided an ideal theoretical scheme, in which the forces are computed from the electronic ground state determined by DFT without any empirical input~\cite{hohenberg1964i,kohn1965self}. However, DFT faces its own challenges. The widely adopted XC functional based on general gradient approximation (GGA)~\cite{Perdew1996b} inherits the self-interaction error and misses the long-range van der Waals (vdW) interaction~\cite{Saek2005,Cohen2012,Cohen2008}. As a result, the liquid structure predicted by GGA-AIMD significantly overestimated the liquid structure~\cite{Zhang2013a,Bankura2013,Bankura2015,DelloStritto2020}. Not surprisingly, the overestimated anion-water interaction again leads to an underestimated energetic of $\rm Cl^-$ anion relative to the solvent as previously reported in literatures~\cite{Ge2013,Zhang2013a,distasio2014,Gaiduk2014,Bankura2015,DelloStritto2020}. Therefore, because of the delicate nature of hydrogen bond (H-bond), the prediction of the aqueous solutions needs a higher level of XC functionals than most of ordinary materials. By mixing a fraction of exact exchange, the vdW inclusive hybrid XC has been shown to soften the liquid structure~\cite{zhang2011s,distasio2014}. Based on the obtained solvation structures~\cite{Gaiduk2014,Bankura2015,Gaiduk2016}, the underestimated energetic of $\rm Cl^-$ anion was found to be largely corrected~\cite{Gaiduk2016}. Furthermore, the hydrogen is the lightest atom, whose nuclear quantum effects (NQEs) cannot be neglected. Under the NQEs, an approximately broadening effect has been reported by the delocalized protons, which slightly softens the water structure~\cite{Tuckerman1997,Morrone2008,Ceriotti2013a,Sun2018a,Li2011}. However, a more important effect has been reported recently~\cite{Xu2020} on the $\rm Cl^-$ solvation structure. Due to the competing NQEs between water-water H-bond and anion-water H-bond, more water molecules, particularly non-bonded water, enter the $\rm Cl^-$ solvation shell. The perturbed solvation structure under NQEs is expected to nontrivially change the local electronic structure. However, such studies remain elusive so far.

\par In this work, we study the NQEs on the electronic structure of solvated chloride anion in water. The equilibrated liquid structure of $\rm Cl^-$ ionic solution is generated by the Feynman path-integral {\it ab initio} molecular dynamics (PI-AIMD) simulation based on the strongly constrained and appropriately normed (SCAN) meta-GGA exchange-correlation approximation~\cite{Sun2015}. With the obtained equilibrated structure, we then compute the quasiparticle energy levels of ionic solutions based on GW approximation~\cite{Onida2002}. In the above, the static limit and frequency dependence on self-energies are both considered by the static-COHSEX and generalized plasmon pole (GPP) models~\cite{Hybertsen1986,Deslippe2012}, respectively. For comparison, the electronic structure calculations are carried out at the PBE-DFT and SCAN-DFT levels as well. Under the NQEs effect, the average distance between anion and water is slightly increased, which increases coordination number particular by the nonbonded water molecules~\cite{Xu2020}. Because of the weaker H-bond between anion and water molecular, the hybridization between $\rm Cl^-$ $3p$ and oxygen $2p$ orbital becomes less strong compared to those in classic AIMD simulations. Therefore, the energy separation between $\rm Cl^-$ level and water band structure is further increased by 0.40 eV towards the experimental direction. Our work shows that the NQEs on solvation structure of anions lead to a small but non-negligible effect on the electronic levels of the hydrated ions. 

\begin{figure*}[htbp]
	\setlength{\abovecaptionskip}{0.cm}
	\includegraphics[width=6.5in]{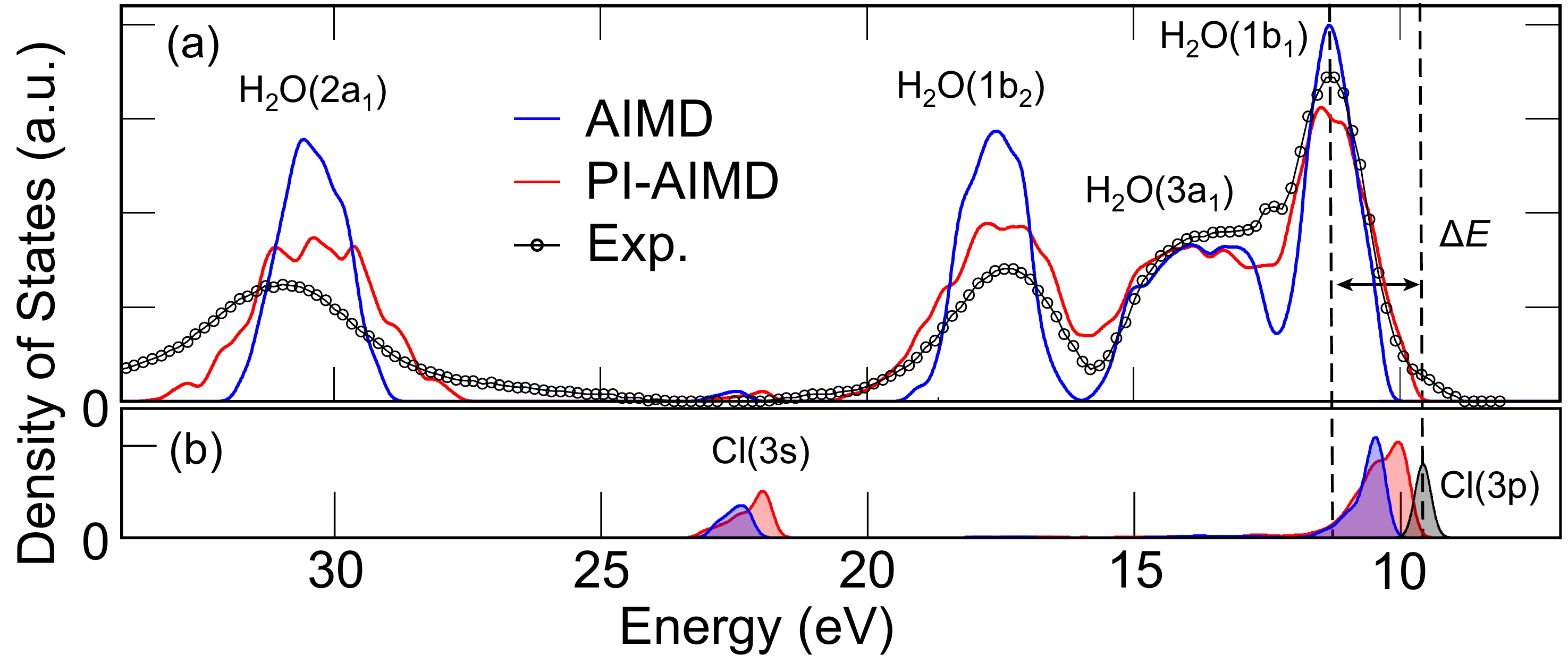}
	\caption{\label{fig:dos}
		The electronic qDOS (a) and $\rm Cl^-$ anion projected qDOS (b) from AIMD (blue) and PI-AIMD (red) trajectories calculated using GW approximation at the GPP level. Experimental data (black circles) is taken from Ref.~\cite{Gaiduk2016}. The peaks are the qDOS projected onto the $\rm Cl^-$ anion orbitals. Note that the experimental position of the qDOS projected onto the $\rm Cl^-$ anion orbital the projected qDOS (Exp.) is extracted from Ref.~\cite{Gaiduk2016} using the spectral difference between ionic solution and neat water in PES experiments. The $\rm Cl^-$ anion-projected qDOS ($3s$ and $3p$ orbitals) have been multiplied by a factor of 5 to be more visible. Both AIMD and PI-AIMD spectra are aligned at $\rm H_{2}O$ ($\rm 1b_1$) peak of the experimental data. All three qDOS spectra are normalized by using the peak area. The qDOS and projected qDOS for $\rm Cl^-$ are averaged over 16 structures of the solvent structure (16 independent structures for AIMD, 16 independent structures from 2 snapshots with 8 bead for PI-AIMD). }
	\vspace{-1.5em}
\end{figure*}

	\vspace{-1.0em}
\section{METHODS}
\label{part2}
\par We simulated a 0.87 M aqueous solution of $\rm Cl^-$ anion using AIMD and PI-AIMD with the SCAN XC functional. We used the simulation trajectory which was reported in Ref.~\cite{Xu2020}. We will briefly explain the simulation details here. All the AIMD and PI-AIMD simulations were performed in the $NVT$ ensemble at $T$ = 300 K with a periodic boundary condition. The cubic cell size is 12.42 $\rm \AA$. One $\rm Cl^-$ anion and 63 $\rm H_{2}O$ water molecules were included. All first-principle molecular dynamics simulations used the Born-Oppenheimer approximation. The potential energy surface used in both PI-AIMD and AIMD simulations were based on the SCAN~\cite{Sun2015} XC functional. The time step for the equation of motion was set to 0.48 fs for both AIMD and PI-AIMD simulations. Total trajectories of $\sim 40$ ps and $\sim 15$ ps were collected after a 5 ps equilibrium for AIMD and PI-AIMD simulations, respectively. All the AIMD and PI-AIMD simulations are carried out using Quantum Espresso~\cite{Giannozzi2017a} and i-PI~\cite{Kapil2019} packages. The $\rm G_{0}W_{0}$ calculations for the energy levels were performed in the static limit (COHSEX) and the GPP levels~\cite{Hybertsen1986} on top of a PBE ground state, where the solvent structure was obtained from either AIMD or PI-AIMD. To compensate the limited number snapshots used in the $\rm G_{0}W_{0}$ calculations, we furthermore computed the energy levels at the DFT level with more snapshots. Our $\rm G_{0}W_{0}$ calculations and DFT calculations were done with the BerkeleyGW~\cite{Hybertsen1986,Deslippe2012} package and Quantum Espresso~\cite{Giannozzi2017a}, respectively. The details of the MD simulation, the GW calculation and DFT calculation could be found in the Supplemental Material~\cite{sm2021}.

	\vspace{-1.0em}
\section{RESULTS and DISCUSSIONS}
\par As shown in Fig.~\ref{fig:dos}(a), we present the computed quasiparticle density of state (qDOS) by GPP-G$_{0}$W$_{0}$ for the valence electrons in $\rm Cl^-$ solution based on equilibrated configurations from both AIMD and PI-AIMD simulations. For comparison, the spectrum obtained in PES experiment in NaCl solution is also shown in the Fig.~\ref{fig:dos}(a), which is aligned with the theoretical predictions at the peak position of electronic states with $\rm 1b_1$ character. 

\par In Fig.~\ref{fig:dos}(a), it can be seen that the overall qDOS in solution is mainly determined by the characteristics of electronic states in water. As a function of increased energies, the four main features in qDOS are associated with $\rm 1b_1$, $\rm 3a_1$, $\rm 1b_2$, and $\rm 2a_1$ orbital symmetries centered at 11.31 eV, 13.78 eV, 17.41 eV, and 30.90 eV in experimental data, respectively. The qDOS computed from ionic solutions generated by classical AIMD simulations predicts rather sharp features compared to experiment~\cite{Xu2020}. By including the NQEs, the above discrepancy is largely corrected by using configurations obtained from PI-AIMD simulations as shown in Fig.~\ref{fig:dos}(a). The quantum nuclei can probe the extended configuration space inaccessible to classical nuclei. Therefore, the protons are significantly more delocalized in both along and normal to the H-bond compared to classical ones~\cite{Ceriotti2013a}.  The proton delocalization in turn produces the observed broadening effect in qDOS, which improves the agreement between experiment and theory in Fig.~\ref{fig:dos}(a)~\cite{Chen2016}. Moreover, comparing qDOS by AIMD and that by PI-AIMD, much larger broadening effects can be identified on the $\rm 1b_2$ and $\rm 2a_1$ orbitals than those on the $\rm 1b_1$ and $\rm 3a_1$ electron.  This is not surprisingly. The bonding pair electrons are mainly comprised of $\rm 1b_2$ and $\rm 2a_1$ states~\cite{Xu2020iso}, therefore, they are more affected by the quantum nuclei. On the other hand, the lone pair electrons, that are mainly constructed by $\rm 1b_1$ and $\rm 3a_1$ orbitals~\cite{Xu2020iso}, are located in the vicinity of the oxygen atom. As a result, they are much less affected by proton displacement under NQEs. 

\par We next focus on the energy levels of solvated $\rm Cl^-$ anion. To this end, we carry out the partial projections of qDOS onto the $\rm Cl^-$ anion for both AIMD and PI-AIMD as shown in Fig.~\ref{fig:dos}(b). For comparison, the experimentally determined $\rm Cl^-$ $3p$ state is also presented in Fig.~\ref{fig:dos}(b), which is extracted from~\cite{Gaiduk2016} by using the spectral difference between ionic solution and neat water in PES experiments. The energy difference $\Delta E_{p}$ between the $3p$ state of the hydrated $\rm Cl^-$ and the $\rm 1b_1$ states of the water has been conveniently used as a signature of the interaction strength between the solute and solvent in energetics~\cite{Zhang2013a,Gaiduk2014,Bankura2015,Gaiduk2016,DelloStritto2020}. Compared to experiment, the predicted $3p$ states of hydrated $\rm Cl^-$ by AIMD is almost submerged in $p$ band of water leading to an underestimated  $\Delta E_{p}$ = 0.85 eV compared to the experimental value of 1.71 eV in Table~\ref{tab:deltaE}. However, by treating the nuclei quantum mechanically in PI-AIMD, the predicted $\Delta E_{p}$ = 1.25 eV is largely increased towards the experimental direction. In order to systematically study this effect, we further carry out electronic structure calculations based on DFT at PBE and SCAN levels, and static G$_{0}$W$_{0}$ as implemented at the COHSEX level, and the results are shown in Table~\ref{tab:deltaE} for both AIMD and PI-AIMD. It can be seen from Table~\ref{tab:deltaE} that $\Delta E_{p}$ computed by PI-AIMD configurations robustly show an increased energy gap between hydrated ion and water. Similar trends can be found in the energy difference $\Delta E_{s}$ between the $3s$ state of the hydrated $\rm Cl^-$ and the $\rm 2a_1$ states of the water, which are shown in Table~\ref{tab:deltaE}. The impact of the NQEs on $s$ band is less prominent due to the fact that $s$ band is deeper than $p$ band. Moreover, the predictions by DFT in general severely underestimate $\Delta E_{p}$ compared to the quasiparticle theories, which is consistent with the previous studies~\cite{Gaiduk2016}. In many spectroscopy experiments associated with electronic excitations in water, a broadening effect in the spectra is often reported in literatures due to the NQEs~\cite{kong2012,Harada2013,wernet2016,Chen2016,Sun2018a}. In the $\rm Cl^-$ ionic solution, the shifted energy distribution of $\rm Cl^-$ electronic states under NQEs is somehow unusual, and it should be attributed to more nontrivial changes in the solvation structure instead of a uniform proton delocalization throughout the liquid.

%%%%%%%%%%%%%%%%%%%%%%%%%%%%%%%%%%%%%%%%%%%%%%%%%%%%%%%%%%%%%%%%%%%%%%%%%%%%%%%%%%%%%%%%%%%%%%%%%%
\begin{figure}[t!]
		\vspace{-0.3em}
	\setlength{\abovecaptionskip}{0.cm}
	\includegraphics[width=3.45in]{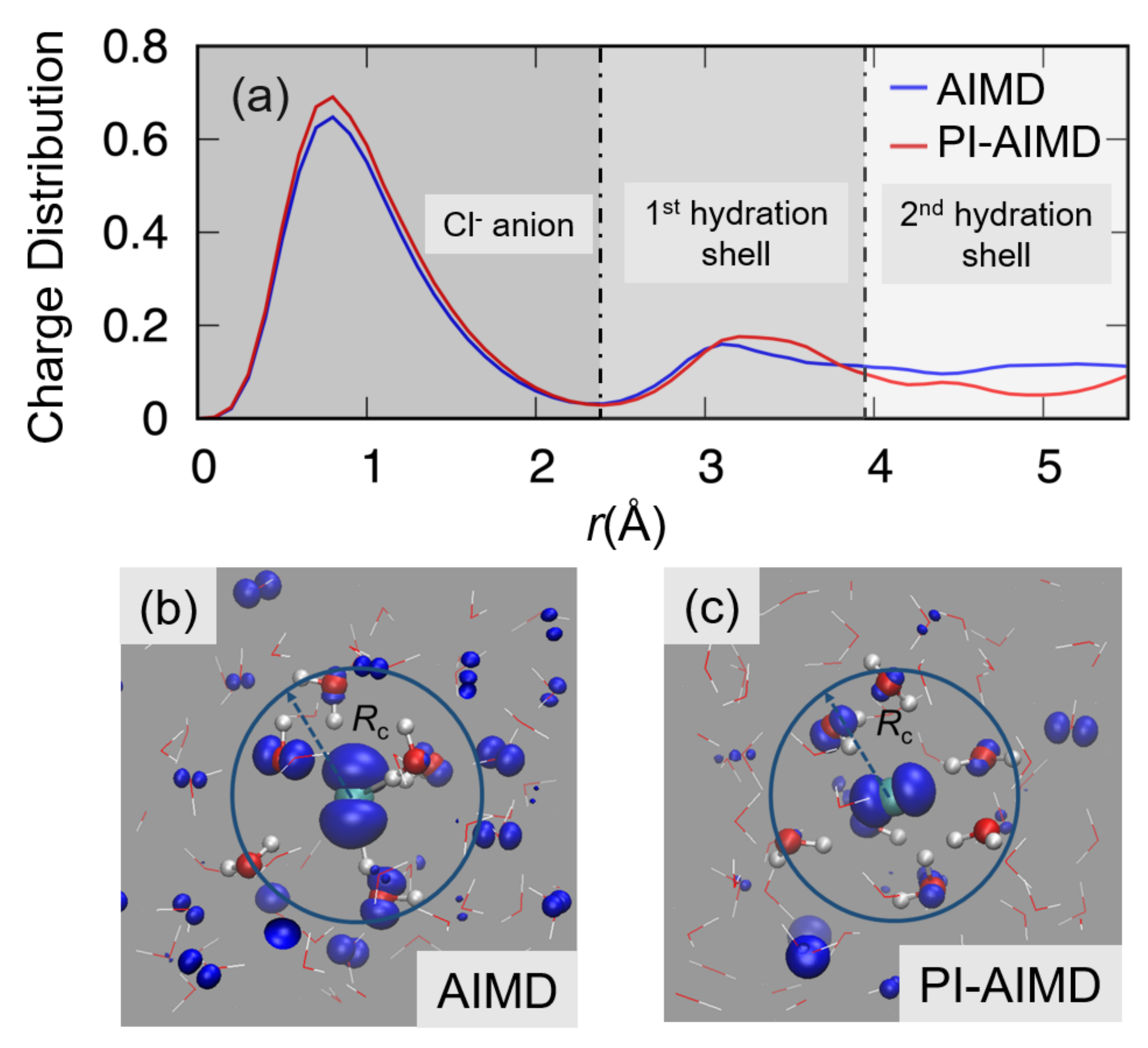}
	\caption{\label{fig:gofr}
		(a) The charge distribution of the $\rm Cl^-$ $3p$ orbital in the AIMD and PI-AIMD where zero is the position of the $\rm Cl^-$ nucleus. The schematic of the spatial distribution of the $\rm Cl^-$ $3p$ orbital hybridization within the anion’s first hydration shell is shown in (b) at the AIMD level and in (c) at the PI-AIMD. $R_c$ is the cutoff distance for the anion’s first hydration shell. The charge density is in blue color. Note that we highlight the water molecules within the first hydration shell and the wavefunction showed here is calculated by using the PBE XC functional, where we sampled total 80 different snapshots to obtain the charge distribution.}
	\vspace{-1.5em}
\end{figure}
%%%%%%%%%%%%%%%%%%%%%%%%%%%%%%%%%%%%%%%%%%%%%%%%%%%%%%%%%%%%%%%%%%%%%%%%%%%%%%%%%%%%%%%%%%%%%%%%%%

\par In the solution, the $\rm Cl^-$ anion is H-bonded to the protonic ends of the water molecules in its first solvation shell. In terms of electronic structure, the above interactions between anion and water molecules are reflected by the degree of hybridization between the $3p$ state of the $\rm Cl^-$ anion and the $p$-band of liquid water. As presented in Fig.~\ref{fig:gofr}(a), we have computed the spatial distribution of the $3p$ state of $\rm Cl^-$ orbital density. With the first hydration shell (from 2.3 $\rm \AA$ to 3.8 $\rm \AA$), Figs.~\ref{fig:gofr} (b) and (c) show that the $3p$ orbital is slightly more localized in the PI-AIMD structure than that in the AIMD. Consistently, in the second hydration shell and beyond (3.8 $\rm \AA$), the density distribution of the $3p$ orbital decays quickly in the PI-AIMD structure that that generated from AIMD trajectory. The more localized $\rm Cl^-$ $3p$ state in the PI-AIMD structure indicates that the hybridization between $3p$ of the anion and $p$-band of water is less favorable in the PI-AIMD structure, which agrees with the larger energy separation between $\rm H_{2}O$ ($\rm 1b_1$) band and $\rm {Cl}^-$ ($3p$) band.

\par
\begin{figure}[!t]
	%	\vspace{0.3em}
	\setlength{\abovecaptionskip}{0 em}
	\includegraphics[width=3.35in]{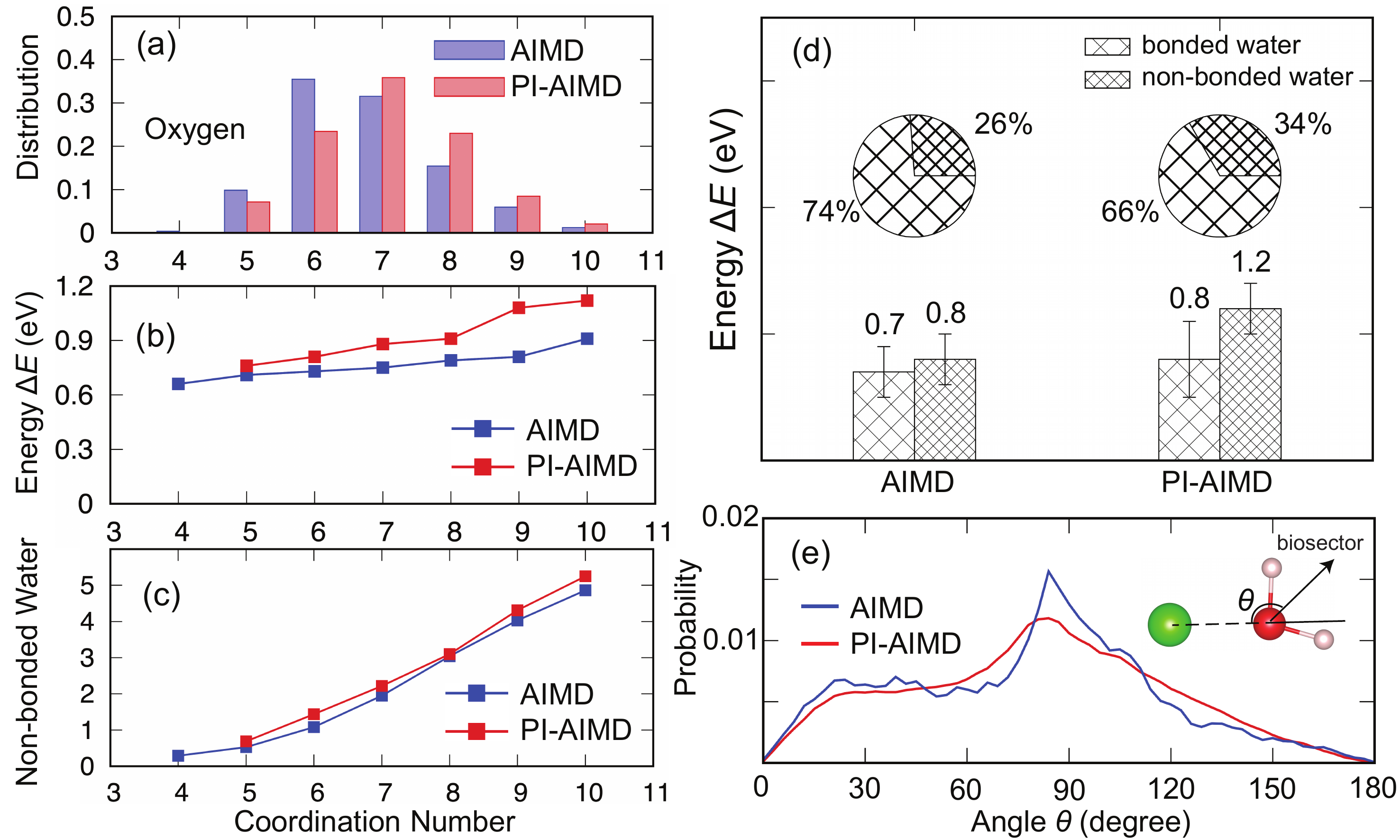}
	\caption{\label{fig:nonbonded}
		(a) The coordination number of oxygen atoms within the first hydration shell of AIMD (blue) and PI-AIMD (red). (b) The averaged energy separation $\Delta E$ with respect to the coordination number of oxygen atoms based on the  AIMD (blue) and PI-AIMD (red). Note that the energy calculation were based on DFT at the level of SCAN. (c) The averaged number of the non-bonded water moleclues within first hydration shell respect to the coordination number of oxygen atoms. (d) The averaged contributions to the energy separation $\Delta E$ of the bonded water and non-bonded water within the first hydration shell. The chart plots are the averaged population of the bonded water and the non-bonded water molecules within the first hydration shell. Note the energy calculation was based on G$_0$W$_0$@GPP. (e) The angle distribution of the non-bonded water within the first hydration shell, where the angle is defined by the Cl-O and the bisector of H-O-H. }
	\vspace{-2 em}
\end{figure}

 The increased energy separation in PI-AIMD compared to that in AIMD, which resulting from the less hybridization of the $\rm Cl^-$ $3p$ electron with water, comes from the fact that the weakened anion-water interaction because of the increased distance between the $\rm Cl^-$ anion and O atoms, as well as the changes of the solution pattern around the polarizable $\rm Cl^-$ anion induced by NQEs. First, we find that the averaged distance of Cl-O in PI-AIMD is 3.34 $\rm \AA$, which is slightly larger than 3.29 $\rm \AA$ in AIMD. This is because that the HB strength of Cl-water pairs is weaker than that of water-water pairs due to the NQEs~\cite{Xu2020}. Second, to show the relation between energy separation and solution pattern, we calculated the coordination number of oxygen atoms shown in Fig.~\ref{fig:nonbonded}(a) within the first hydration shell, where it is defined as when the Cl-O distance is smaller than 3.82 $\rm \AA$ and 3.84 $\rm \AA$ for AIMD and PI-AIMD, respectively. We find that under the influence of NQEs, the averaged coordination number in PI-AIMD (7.08) is larger than that in the AIMD (6.62). This is consistent with the increased Cl-O distance in PI-AIMD, since more space will be allowed for the surrounding water molecules. We decompose the energy separation contribution in terms of the coordination number, shown in Fig.~\ref{fig:nonbonded}(b). One can find two trends: First, the energy separation gradually increase with respect to the coordination number for either AIMD or PI-AIMD. It is obvious that with more water molecules within the first hydration shell, the Cl-O distance will increase, leading to the increase of the energy separation. Second, the energy separations in PI-AIMD are larger than that in AIMD for each coordination number, indicating some nontrivial structural changes. 
\par The polarizable $\rm Cl^-$ anion makes the surrounding water molecules tend to populate one side of the anion and leave the other side relatively empty in the solution~\cite{Ohtaki1993,Tobias2001,Ge2013,DelloStritto2020,Xu2020,Migliorati2014,Perera1992}. The abilities to polarize water are different for $\rm Cl^-$ anion-water H-bond and water-water H-bond, where $\rm Cl^-$ anion-water H-bond has a weaker bonding strength than that of water-water H-bond under the influence of the NQEs~\cite{Xu2020}. We decompose the water molecules which are within the first hydration shell of $\rm Cl^-$ into two categories, one is bonded with $\rm Cl^-$ anion, another one is non-bonded with $\rm Cl^-$ anion. Note that if the Cl-O distance of the water molecule is smaller than 3.9 $\rm \AA$ and Cl-O-H angle is smaller than 30$^\circ$, this water molecule will be counted as bonded to the $\rm Cl^-$ anion~\cite{Wang2013a}. We compute the changes of the number of non-bonded water with respect to the coordination number, which is shown in Fig.~\ref{fig:nonbonded}(c). One can find that the number of non-bonded water molecules gradually increase, while the number of bonded water gradually become saturated. Moreover, we find that the average number of bonded water molecules within the first hydration shell of the $\rm Cl^-$ anion decreases from 4.89 in the AIMD structure to 4.71 in the PI-AIMD structure, while the average number of the non-bonded water molecules increases from 1.73 in AIMD to 2.37 in PI-AIMD, the fractions are shown in the inserted chart of Fig.~\ref{fig:nonbonded}(d).

%%%%%%%%%%%%%%%%%%%%%%%%%%%%%%%%%%%%%%%%%%%%%%%%%%%%%%%%%%%%%%%%%%%%%%%%%%%%%%%%%%%%%%%%%%%%%%%%%%
%\begin{figure}[!htb]
%%%%%%%%%%%%%%%%%%%%%%%%%%%%%%%%%%%%%%%%%%%%%%%%%%%%%%%%%%%%%%%%%%%%%%%%%%%%%%%%%%%%%%%%%%%%%%%%%%

\par To reveal the relation between energy separation and geometry of the bonded/non-bonded water molecules, we compute the energy separation between the projected qDOS peak positions of the $\rm Cl^-$ anion and the $\rm H_{2}O$ ($\rm 1b_1$) orbitals coming from water molecules bonded to the $\rm Cl^-$ anion, $\rm \Delta E_{bonded} = E(H_{2}O_{bonded}(1b_1)) - E(Cl^-(3{\it p}))$, verses non-bonded water molecules, $\rm \Delta E_{non-bonded} = E(H_{2}O_{non-bonded}(1b_1)) - E(Cl^-(3{\it p}))$. Both calculations consider only water molecules within the first hydration shell of $\rm Cl^-$ anion. The data are shown in Fig.~\ref{fig:nonbonded}(d). We find that on average, in the AIMD case, $\rm \Delta E_{non-bonded} = 0.8 \pm 0.3$ eV and $\rm \Delta E_{bonded} = 0.7 \pm 0.2$ eV; however, under the influence of NQEs, the energy separation between the non-bonded water and $\rm Cl^-$ anion increases to $\rm \Delta E_{non-bonded} = 1.2 \pm 0.3$ eV, while $\rm \Delta E_{bonded} = 0.8 \pm 0.3$ eV. The energy difference of the non-bonded water in PI-AIMD and AIMD could be explained by their geometry difference. We compute the angle distribution of the non-bonded water within the first hydration shell, where the angle is defined by the Cl-O and biosector of H-O-H, shown in Fig.~\ref{fig:nonbonded}(e). One can find that the angle predicted by PI-AIMD is larger than that predicted by AIMD, leading to averaged angles $\sim85^\circ$ for PI-AIMD and $\sim78^\circ$ for AIMD, respectively. This indicates that under the influence of NQEs, non-bonded water molecules tend to point away from $\rm Cl^-$ anion. Moreover, we find that the averaged distance between $\rm Cl^-$ anion and O atoms for the non-bonded water molecule is slightly increased from 3.45 $\rm \AA$ (AIMD) to 3.55 $\rm \AA$ (PI-AIMD). These geometry changes are the indicators of the weakened interaction between the $\rm Cl^-$ anion and the non-bonded water molecules. 

\par In short, the origin of the increased energy separation between the $3p$ state of hydrated $\rm Cl^-$ and the $\rm 1b_1$ states of liquid water comes from the following aspects: First, the averaged energy separation for the non-bonded water molecules and $\rm Cl^-$ anion in the PI-AIMD simulation is larger than the one in the AIMD simulation due to the larger distance and pointing away non-bonded water molecules induced by NQEs; Second, the fraction of non-bonded water molecules within the first hydration shell increases in the PI-AIMD simulation comparing to that in the AIMD simulation, which indicates that more water molecules are able to bond with other water molecules instead of the $\rm Cl^-$ anion. All these geometry changes lead to less hybridization of the $3p$ orbital of the $\rm Cl^-$ anion with the water molecules and increases the energy separation between the $\rm Cl^-$ ($3p$) band and the $\rm H_{2}O$ ($\rm 1b_1$) band.

\vspace{-1.5em}
\section{CONCLUSION}
\par In summary, we have investigated the NQEs on the electronic structure of the hydrated $\rm Cl^-$ anion based on the frequency dependent G$_0$W$_0$ quasiparticle computational approach. With the NQEs considered in molecular structure, the interaction between the anion and water is effectively weakened. As a result, the coordination number in the first hydration shell increases, which is particularly accompanied by an increased population of interstitial water molecules. We have shown that the above weaker anion-water interaction also suppresses the electronic hybridization between $\rm Cl^-$ and water, which in turn increases the energy gap between $3p$ state of $\rm Cl^-$ and the $2p$ state of water. The above computed energy gap of 1.25 eV shows a better agreement with PES experimental value of 1.71 eV. In contrast, the energy gap computed from trajectories by classic molecular dynamics simulations yields a largely underestimated value of 0.85 eV.
\par In the last decade, continuous efforts have been devoted to understanding the interaction between chloride anion and surrounding water focusing on the electronic structure that can be probed by the PES experiments. In addition to the corrections previously demonstrated by the hybrid density functional and vdW inclusive AIMD simulations~\cite{Bankura2015,Gaiduk2016}, our current work adds an important physical effect which further improve the agreement between theory and experiment by quantum nuclei. It is noted that our current theory still slightly underestimates the energy separation between the anion and water. Since the SCAN functional also inherits the self-interaction error, it is expected that the remaining discrepancy should be further corrected by the quasiparticle calculations from the hybrid DFT based PI-AIMD trajectory with less self-interaction error. At the same time, improvement towards experiment is also expected by using better starting wavefunctions in the G$_0$W$_0$ perturbation theory.

	\vspace{-0.5em}
\begin{acknowledgments}
\par We thank Dr. Zhenglu Li for helpful discussions. This work was primarily supported by the Computational Chemical Center: Chemistry in Solution and at Interfaces funded by The DoE under Award No. DE-SC0019394 (F. T. and X. W.). D.Y.Q. was supported by the National Science Foundation (NSF) under grant number DMR-2114081. Part of the computational work was performed at the Molecular Foundry, which is supported by the Office of Science, Office of Basic Energy Sciences, of the U.S. Department of Energy under Contract No. DE-AC02-05CH11231. This research used resources of the National Energy Research Scientific Computing Center (NERSC), a U.S. Department of Energy Office of Science User Facility located at Lawrence Berkeley National Laboratory, operated under Contract No. DE-AC02-05CH11231. This research includes calculations carried out on HPC resources supported in part by the National Science Foundation through major research instrumentation grant number 1625061 and by the US Army Research Laboratory under contract number W911NF-16-2-0189.
\end{acknowledgments}

\bibliography{chloride_ion}

\end{document}